\documentclass[twocolumn,superscriptaddress,secnumarabic,amssymb, nobibnotes, aps, prl]{revtex4-2}
\usepackage{graphicx}
\usepackage{amsfonts}
\usepackage{amssymb}
\usepackage{amsmath}
\usepackage{psfrag}
\usepackage{natbib}
\usepackage{mathtools}
\usepackage[dvipsnames]{xcolor}

\usepackage{ulem}
\usepackage{float}
\usepackage[colorlinks=true, linkcolor=blue, urlcolor=blue, citecolor=black]{hyperref}
\usepackage{caption}
\newcommand{\beqa}{\begin{eqnarray}}
\newcommand{\eeqa}{\end{eqnarray}}
\newcommand{\beq}{\begin{equation}}
\newcommand{\eeq}{\end{equation}}

\usepackage[justification=justified,singlelinecheck=false]{caption}

\begin{document}

\title{Peeling-Induced Rolling and Heterogeneous Adhesion in Blistered Films}

\author{Amit Kumar Pandey}
 \affiliation{Mechanical \& Aerospace Engineering Department, Syracuse University, Syracuse, NY 13244, USA}
\author{Pei Ren-Sawyer}
 \affiliation{Mechanical \& Aerospace Engineering Department, Syracuse University, Syracuse, NY 13244, USA}
\author{Sunghwan Jung}
 \affiliation{Department of Biological \& Environmental Engineering, Cornell University, Ithaca, NY 14853, USA}
\author{Teng Zhang}
 \affiliation{Mechanical \& Aerospace Engineering Department, Syracuse University, Syracuse, NY 13244, USA}
  \affiliation{BioInspired Syracuse: Institute for Material and Living Systems, Syracuse University, Syracuse, NY 13244, USA}
\author{Anupam Pandey}
\email{apande05@syr.edu}
 \affiliation{Mechanical \& Aerospace Engineering Department, Syracuse University, Syracuse, NY 13244, USA}
 \affiliation{BioInspired Syracuse: Institute for Material and Living Systems, Syracuse University, Syracuse, NY 13244, USA}


\begin{abstract}
            \noindent Blisters, delaminated regions that form in multilayered structures under compressive stresses, are observed across a wide range of length scales, from two-dimensional materials to protective coatings and laminated composites. Far from being passive defects, such interfacial features have emerged as functional motifs for three-dimensional architectures and reconfigurable surfaces. Here we reveal an unusual peel response of a blistered thin film on a soft substrate. When peeled from one end, the advancing peel front triggers reattachment at the blister edge once a critical separation is reached, initiating spontaneous rolling of the film on the substrate. This peel-to-roll transition produces a sharp drop in the measured adhesion force, which then remains constant throughout the rolling phase. Using experiments, scaling analysis, and molecular dynamics simulations, we resolve the contact morphology at the transition and identify the contact length at which rolling initiates. We show that this length arises from interactions between the two contact edges and is independent of the work of adhesion. Once rolling begins, a dynamically imposed dwell time -- defined by the rolling length and peel speed -- translates contact history into spatial variations in adhesion force, thereby governing the magnitude of the force drop. Together, these results point to a new pathway for generating spatially tunable, heterogeneous adhesion from otherwise homogeneous interfaces.
  
\end{abstract}

\maketitle


Removing stickers or tapes from surfaces provides an intuitive sense of the strength of adhesion holding them together. This commonplace experience illustrates the mechanical process of peeling, a versatile method that underpins a broad range of technological operations, from precision tasks such as transfer printing~\cite{meitl2006transfer,jeong2014high} and exfoliation of electronic structures~\cite{yi2015review, choi2016recent} to everyday applications like adhesive tapes and protective coatings~\cite{gu2016understanding,villey2015rate,zhou2021thermal}. Peeling translates complex microscopic interfacial interactions into macroscopic forces, making it a powerful metrological tool for quantifying adhesion~\cite{bartlett2023peel}. While peeling homogeneous interfaces typically yields a constant adhesion force, introducing engineered defects such as cuts~\cite{xia2012toughening, xia2013adhesion}, kirigami structures~\cite{hwang2018kirigami, hwang2023metamaterial}, patterned stiffness~\cite{gao2023toughening, zhu2023mechanics}, or controlled alignment in liquid crystalline elastomers~\cite{pranda2024directional} offers novel pathways to actively tune and spatially modulate adhesion properties. 

One of the most common patterns that naturally arises due to local adhesion failure in multilayer systems is the blister - localized, out-of-plane delamination~\cite{freund2004thin, hutchinson1992growth, gioia1997delamination, audoly1999stability}. Blisters are observed across a wide range of systems, from nanoscale van der Waals heterostructures, prefabricated circuit boards to large-scale laminated composites~\cite{dai2020mechanics, ni2017shape, gille1984buckling}. The onset of blistering and blister morphology are determined by the interfacial adhesion as well as the geometric and material parameters of the film-substrate assembly~\cite{vella2009macroscopic, box2023delamination, mei2007buckling, zhang2018spontaneous, cai2011periodic}. The ability to control blister formation has opened novel strategies for self-assembling three-dimensional architectures and reconfigurable surfaces~\cite{xu2015assembly, liu2019harnessing, ko2009curvilinear}. In addition, controlled blistering has emerged as a powerful metrological tool for probing interfacial strength of ultrathin materials~\cite{cao2014blister, sanchez2018mechanics, cordill2007adhesion}. Despite extensive investigations into the form and function of blisteres, the mechanical response of a blistered film to external forcing, remains poorly understood.

Here we show that instead of serving as a passive defect, a blister fundamentally alters the propagation of interfacial fracture. As an advancing peel front approaches a blister, the film simultaneously detaches at the peel front and reattaches at the blister edge, initiating a rolling motion in which the blister progressively shrinks. This peel-to-roll transition produces a discontinuous force response, selects a finite rolling contact length, and introduces an emergent timescale that cannot be inferred from classical peeling or blistering theories alone. By combining experiments, simulations, and scaling analysis, we capture the onset of rolling, characterize its dynamics and impact on the adhesion jump, and demonstrate how blisters can be harnessed to modulate adhesion in otherwise homogeneous interfaces.

\begin{figure*}[t]
\centering
\includegraphics[width=1\textwidth]{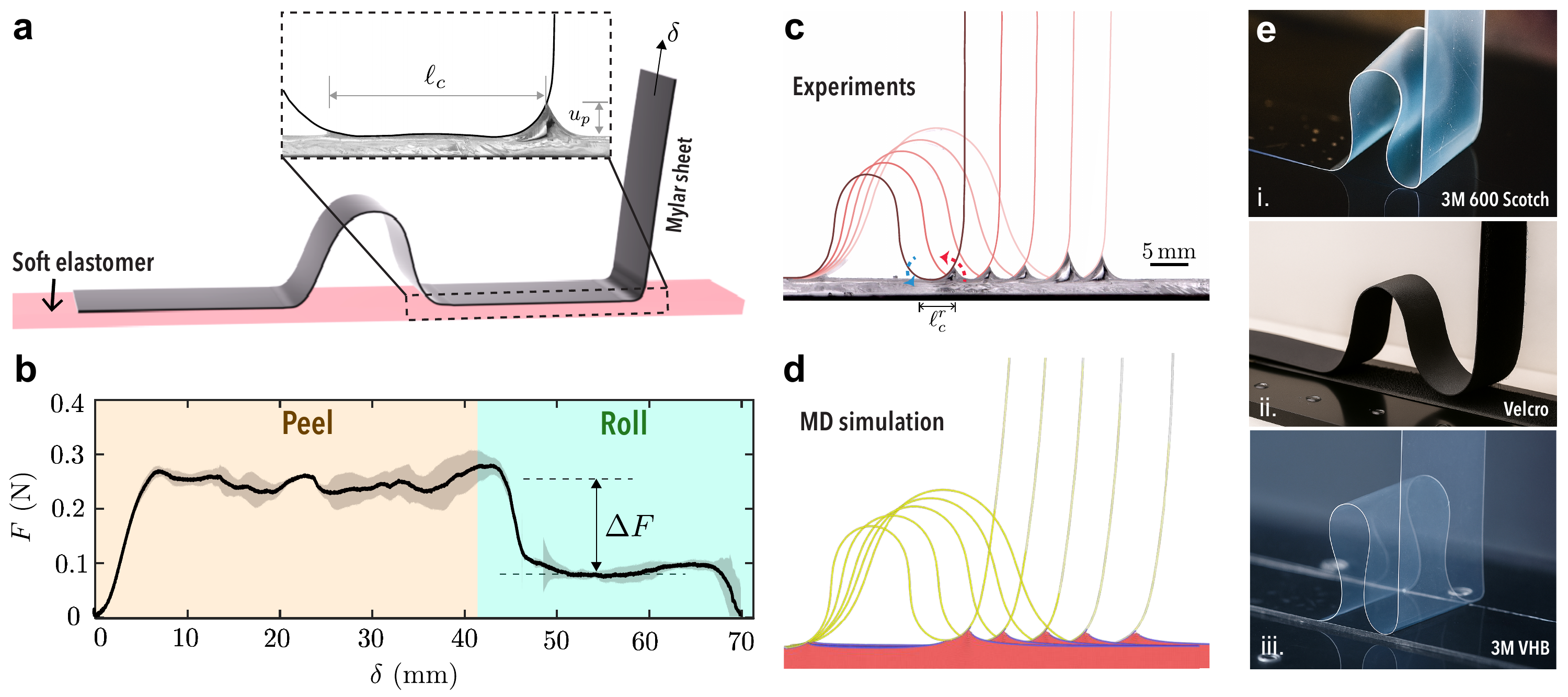}
\caption{\textbf{Peel-to-roll transition and the associated force drop during peeling of sticky blisters}. \textbf{a}. Schematic of a thin Mylar film adhered to a soft silicone elastomer and peeled at a 90$^{\circ}$ angle. As the peel front approaches the blister edge, the contact length $\ell_c$ decreases. Inset: Magnified contact zone showing asymmetric substrate deformation and film curvature. \textbf{b}. Force–displacement curve exhibiting two regimes: steady-state peeling followed by a sudden force drop ($\sim$ 260\%) during rolling. Notably, the drop in force lags behind the onset of rolling. This data corresponds to 25 $\mu$m thick Mylar and 12 mm thick substrate. \textbf{c}. Composite time-lapse image showing the evolution of the blister during rolling. Color gradient from light to dark tracks the progression in time. Arrows mark simultaneous detachment and reattachment, defining a stable rolling contact length $\ell_c^r$. \textbf{d}. Molecular dynamics simulations reproduce the peel-to-roll transition; snapshots are shown for every 1.8 mm of peel displacement. Yellow indicates the film, pink the substrate interior, and blue the substrate surface. \textbf{e}. Robustness across commercial adhesives: (i) Scotch 3M 600 tape on acrylic, (ii) Velcro peeled from itself, and (iii) a blistered mylar sheet peeled from 3M VHB acrylic foam adhesive.}
\label{fig1}
\vspace{-3mm}
\end{figure*}

\vspace{-5mm}
\section{Results}
\vspace{-4mm}
\noindent\textbf{Phenomenology of peeling a blistered film from a soft substrate -}
Our experiments involve laying a buckled thin film on an unstressed silicone elastomer substrate to form a single blister (Fig.~\ref{fig1}a). We use four types of films, mylar, fluorinated ethylene propylene (FEP), optically clear polyester (PET), and transparent perfluoroalkoxy (PFA), with thicknesses ranging from 25.4 to 127 $\mu$m, spanning bending moduli $B$ from $8.74\times 10^{-7}$ to $2.27\times 10^{-4}$ J. The substrate is prepared from Dow Corning CY52-276, mixed in a 1:1 mass ratio and cured in molds of thickness $h_a$ = 1\text{–}12\,\text{mm}. The resulting crosslinked elastomer has a Young’s modulus of 2190 Pa and forms a compliant adhesive layer beneath the film. Peeling is performed from one end of the film at a constant speed and at a 90$^{\circ}$ angle. An Instron 34SC-1 load frame equipped with a 100 N load cell controls the peel speed $V$ and records the peel force $F$ as a function of displacement $\delta$ (see Materials \& Methods for more details).

As the thin film is peeled from the soft substrate, the adhered region ahead of the blister -- of length $\ell_c$ (Fig.~\ref{fig1}a inset) gradually decreases. During this stage, peeling proceeds with a clean interfacial fracture and the measured force remains approximately constant, defining the peel regime in Fig.~\ref{fig1}b. The corresponding work of adhesion, $\Gamma=F/b\simeq 10$ J/$m^2$, far exceeds the surface energy of the elastomer (typically 20-50 mJ/$m^2$)~\cite{xu2017direct,mondal2015estimation}, indicating that the effective or apparent adhesion is dominated by substrate strain energy and viscoelastic dissipation~\cite{creton2016fracture}. When $\ell_c$ falls below a critical value, the system departs abruptly from the classical peeling: the blister starts to shrink. This shrinkage arises from a coupled two-front motion in which the film simultaneously detaches at the peel front and reattaches at the blister edge, producing a rolling-like interfacial motion (cf. SI Movie 1). Notably, the observed rolling behavior persists at lower peel angles (SI Fig. S3) and over a wide range of peel speeds ($10^{-3}-1$ mm/s), indicating that it is not sensitive to a specific peel geometry or rate. The peel-to-roll transition leaves a distinct mechanical signature: a sharp, multifold drop in the force, which remains steady throughout the roll phase (Fig.\ref{fig1}b). The composite time-lapse images in Fig.~\ref{fig1}c show the shapes during rolling: the blister edge propagates inward, selecting a finite rolling contact length $\ell_c^{r}$. The force eventually drops to zero once the blister opens up. We want to point out that in Fig.\ref{fig1}b, the onset of rolling is marked slightly before the force drops to its lower plateau — a subtle but robust feature that we return to in the next section. We corroborate the experiments with a coarse-grained, plane-strain simulation of peeling and rolling using LAMMPS~\cite{thompson2022lammps}. The substrate is modeled as a nonlinear elastic lattice derived from finite element discretization~\cite{zhang2019deriving}, the film as a discrete beam with prescribed bending rigidity, and adhesion through short-range Lennard–Jones interactions (see SI Sec 4 for details). Under quasi-static peeling, simulation reproduces the key experimental response: once the contact length falls below $\ell_c^r$, rolling initiates and the blister shrinks progressively (Fig.~\ref{fig1}d). 

\begin{figure}[t]
\centering
\includegraphics[width=0.48\textwidth]{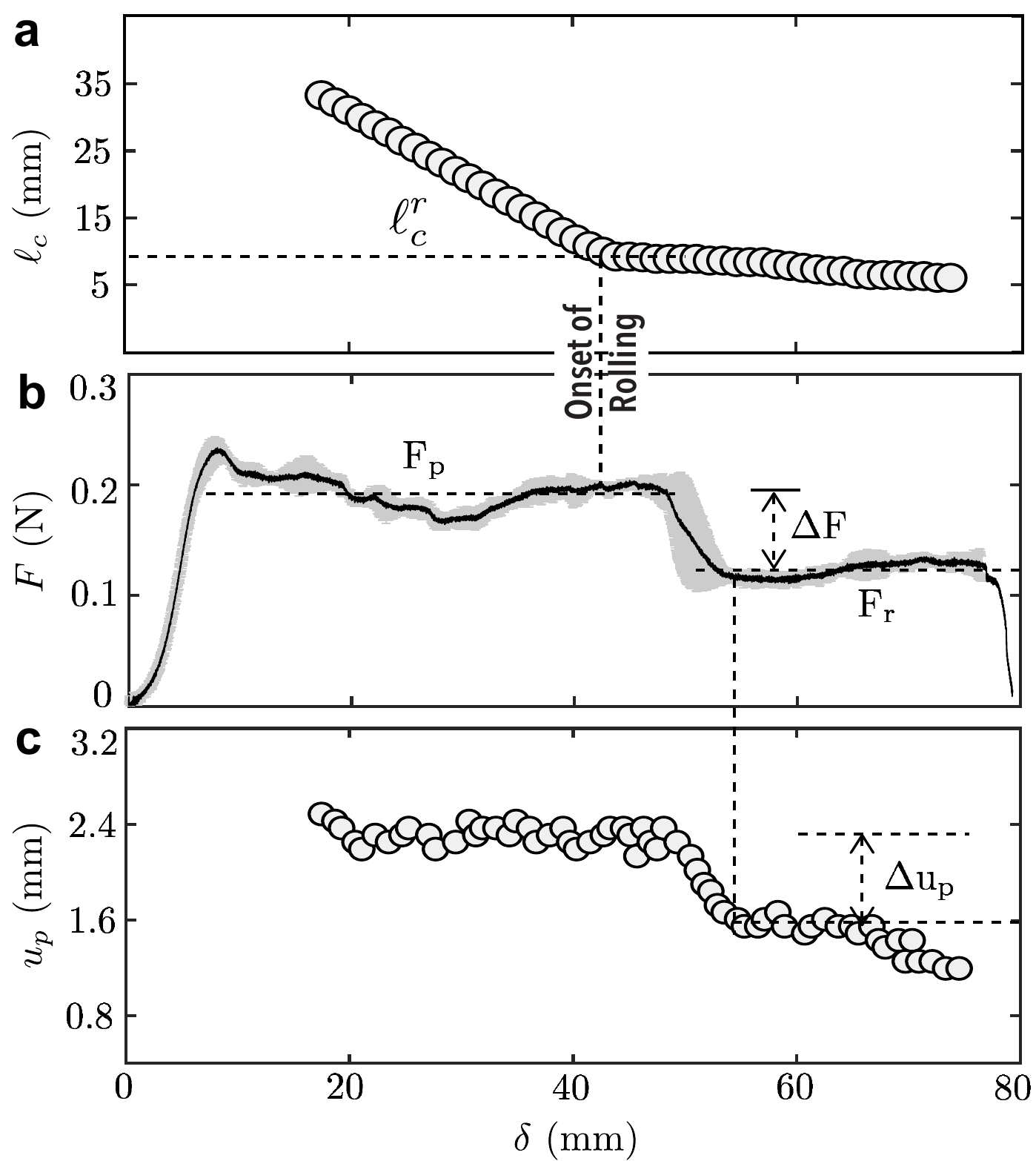}
\caption{\textbf{Contact morphology and delayed force response during the peel‑to‑roll transition.} \textbf{a}. Evolution of contact length $\ell_c$ as the film is peeled. $\ell_c$ decreases steadily during classical peeling until the onset of rolling, at which point it stabilizes at a constant value $\ell_c^r$. \textbf{b}. Adhesion force measured simultaneously shows a delayed response: force remains at the peel value at the onset of rolling, and only drops after an additional $\simeq 10$ mm of peel displacement. \textbf{c}. Substrate deformation at the peel edge, $u_p$, mirrors the force response: it remains large during peeling and decreases sharply once rolling is fully established. These data corresponds to 50 $\mu$m mylar film being peeled from 3.5 mm thick substrate at 0.17 mm/s. }
\label{fig2}
\vspace{-5mm}
\end{figure}

To assess the robustness of this behavior across adhesive systems, we examined a range of commercial tapes that span widely different geometries and adhesion mechanisms: a thin pressure-sensitive adhesive layer (3M Scotch 600), a purely mechanical hook–loop interface with no glue (Velcro), and a backing peeled from a thick compliant acrylic-foam adhesive (3M VHB). Remarkably, all of these systems display the peel-to-roll transition as shown in Fig.~\ref{fig1}e and SI Movie 2. Figure S5 in the SI presents the measured force-displacement responses for eight different commercial tapes, all of which display a sudden force drop associated with the onset of rolling. The consistency of this response -- despite large differences in adhesive strength and the microscopic origin of adhesion -- shows that the onset of rolling is controlled by geometric and mechanical coupling in the system, not by the details of interfacial chemistry.\\

\begin{figure}[t]
\centering
\includegraphics[width=0.49\textwidth]{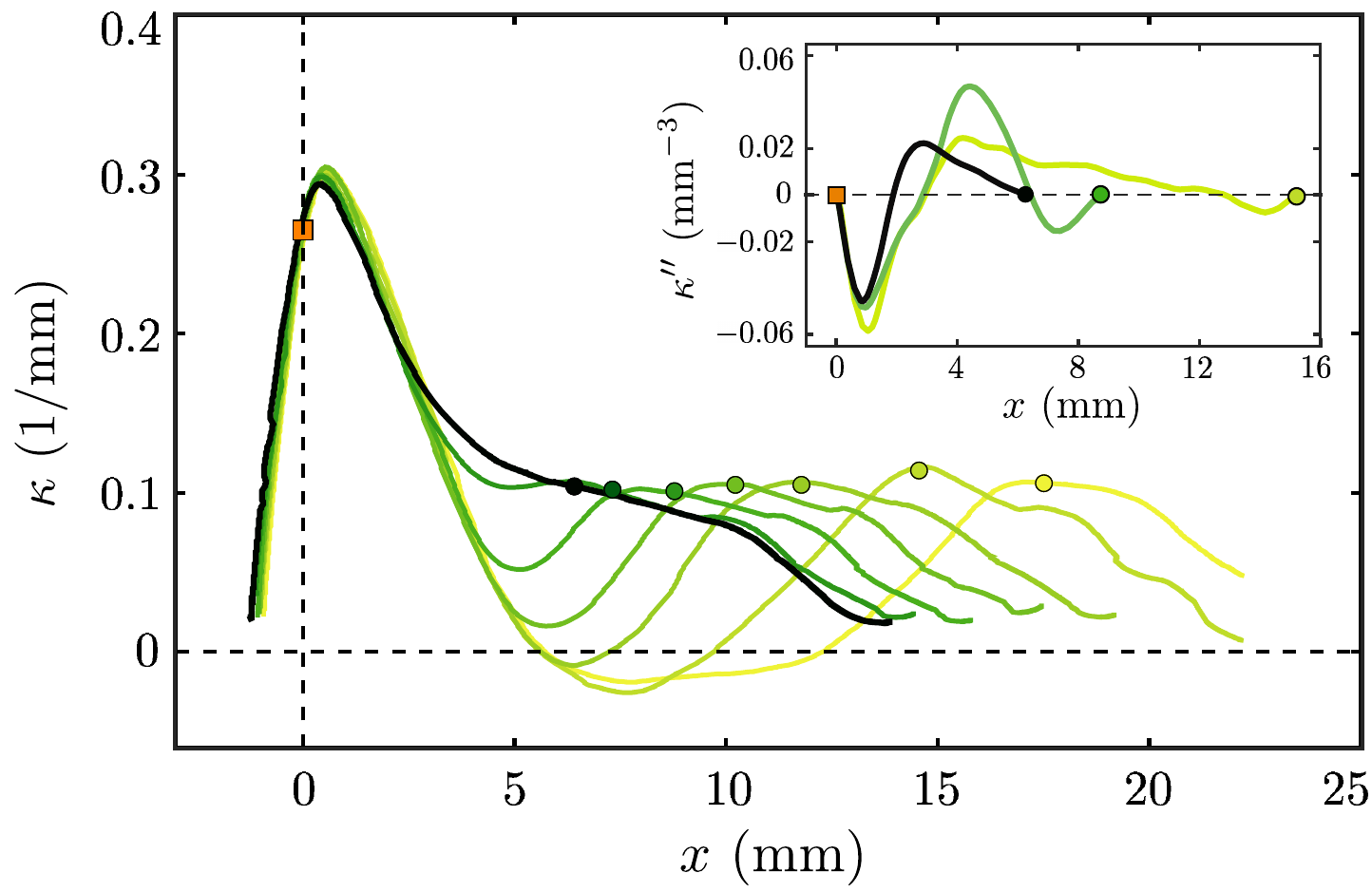}
\caption{\textbf{Curvature evolution within the adhesive contact.} The film curvature $\kappa(x)$ along the adhered region evolves as the contact length $\ell_c$ decreases from the peel phase (yellow) to the onset of rolling (black).The peel front is stationary at $x=0$ (orange square), while blister edge positions at successive times are marked by circles. Inset: $\kappa''$ provides an estimate of the normal traction transmitted to the substrate within the contact. The tensile region at the blister edge disappears at the rolling onset, consistent with the loss of a distinct curvature boundary condition there.}
\label{fig3}
\vspace{-5mm}
\end{figure}

\noindent\textbf{Onset of rolling and the discontinuous force response - }
To relate the evolution of the contact geometry to the measured peel force, we track the contact length $\ell_c$, the peel force $F$, and the substrate deformation at the peel edge $u_p$ as functions of peel displacement $\delta$ (Fig.~\ref{fig2}). Figure~\ref{fig2}a shows that $\ell_c$ decreases steadily during peeling, but undergoes a sharp transition to a constant value $\ell_c^r$ once rolling begins. Notably, this transition occurs before the measured force drops, with the two events separated by a displacement of order $\ell_c^r$. In contrast, the peel-edge deformation $u_p$ decreases in magnitude at the same displacement at which the force drops, reflecting its sensitivity to the strength of the adhesive contact at the peel front. These observations reveal two key features. First, the critical contact length $\ell_c^r\simeq 10$ mm is far larger than the adhesive or film thicknesses -- and remains approximately constant across the range of peel speeds tested (see SI Fig. S6), indicating that it is set by the mechanics of the film–substrate interaction. Second, the delayed force drop shows that rolling itself carries little energetic cost: the film can reattach at the blister edge without an immediate change in force. Instead, the force drop occurs only when the newly formed contacts, created at the blister edge and transported through the adhered region by rolling, are subsequently detached at the peel front. This emergent dwell time or contact age governs the magnitude of the force drop $\Delta F$ and the corresponding change in substrate deformation at the peel front ($\Delta u_p$).

The emergence of a finite $\ell_c^r$ reflects how stress fields within the contact region evolve as the peel front approaches the blister. Both the peel and blister edges impose tensile stresses on the substrate, but with markedly different curvature boundary conditions as shown in Fig.\ref{fig3}: a sharply peaked curvature at the peel edge and a much weaker curvature at the blister edge. For large $\ell_c$, the associated curvature fields relax independently into the contact, producing a pronounced valley in $\kappa(x)$. As $\ell_c$ decreases, the peel-edge curvature field extends deeper into the contact, progressively narrowing this valley. Mechanical equilibrium links film bending to substrate traction through, $p(x)\simeq B\kappa''(x)$, so $\kappa''(x)$, plotted in the inset of Fig.~\ref{fig3} provides an estimate of the pressure transmitted to substrate. As $\ell_c$ approaches $\ell_c^r$, the tensile stress at the blister edge disappears, indicating that compressive stresses generated at the peel edge overwhelm the blister-imposed boundary condition. At this point, the valley in curvature disappears: the blister can no longer sustain its geometry, and the film rotates and reattaches -- marking the onset of rolling.

\begin{figure}[t]
\centering
\includegraphics[width=0.49\textwidth]{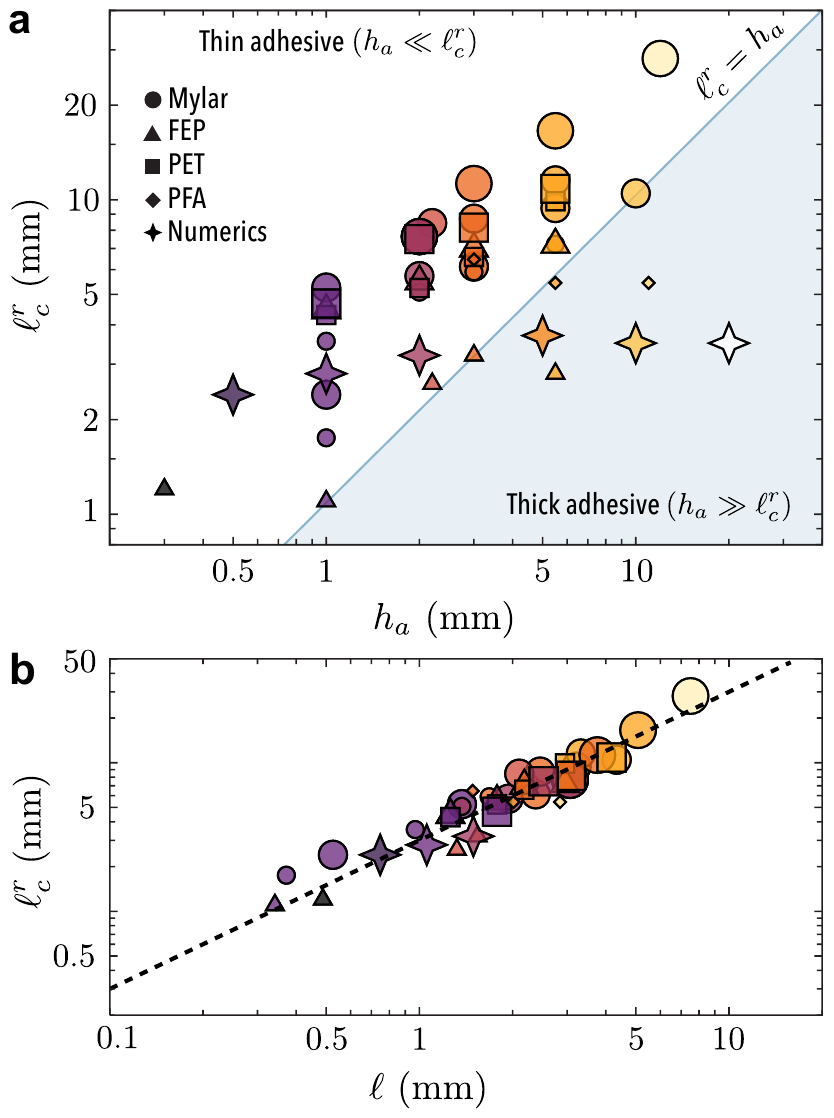}
\caption{\textbf{Contact length at rolling onset}. \textbf{a}. Contact length at the rolling onset are plotted as a function of adhesive thickness ($h_a$). Marker shape denotes film material (Mylar, FEP, PET, PFA), marker size increases with bending modulus $B$, and color gradient from dark to light indicates increasing adhesive thickness. The solid line $\ell_c^r = h_a$ separates the thin-adhesive ($h_a \ll \ell_c^r$) and thick-adhesive ($h_a \gg \ell_c^r$) regimes. Numerical data points corresponds to a 50 $\mu$m thick film with elastic modulus 1 GPa adhered to a soft substrate with modulus 10 kPa. \textbf{b}. $\ell_c^r$ plotted against the predicted decay length $\ell$ computed from Eq.~\eqref{eq3} for data in the thin-adhesive regime ($h_a < \ell_c^r$). The dashed line represents a fit to the collapsed data, yielding $\ell_c^r\simeq3.0\ell$.}
\label{fig4}
\vspace{-3mm}
\end{figure}

\begin{figure*}[t]
\centering
\includegraphics[width=1\textwidth]{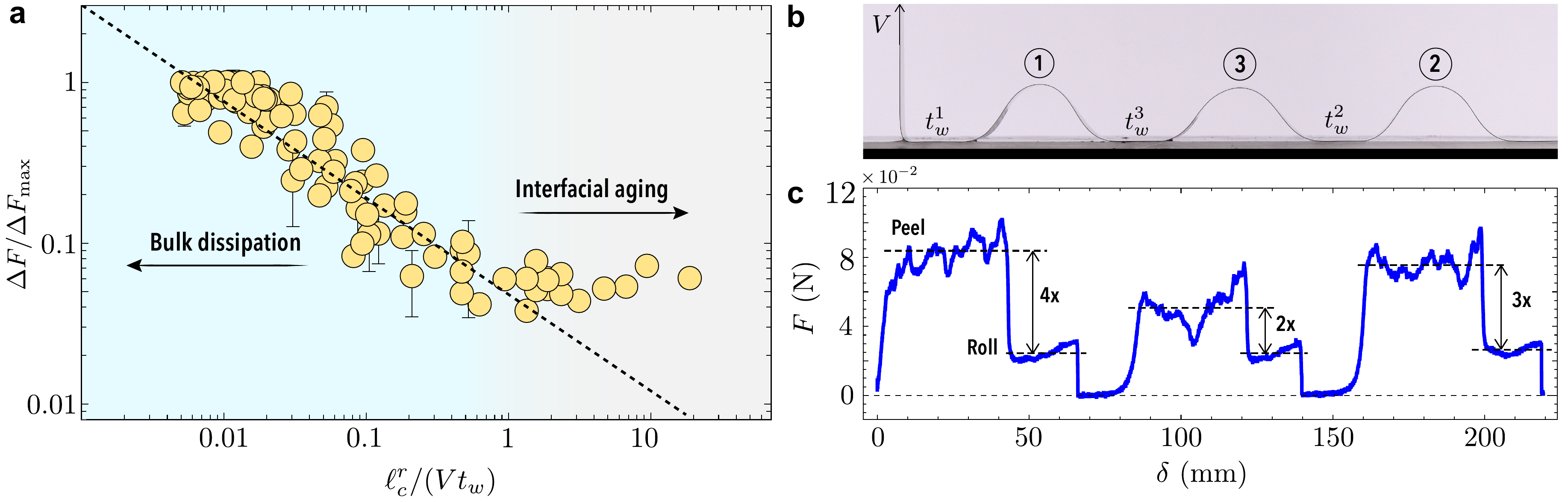}
\caption{\textbf{Force jump at the peel-to-roll transition and multi-step, heterogeneous adhesion landscape in blister array} \textbf{a}. Dimensionless force drop at the transition, ($\Delta F/\Delta F_m$), plotted against the ratio of dwell times in the roll ($\ell_c^r/V$) and peel ($t_w$) phases. Data points correspond to measurements conducted at various peel speeds ($V$), peel phase contact times ($t_w$), and $\ell_c^r$ values. The dashed line represents a power-law with exponent matching the elastomer’s rheological exponent $n=0.55$. For $\ell_c^r/(V t_w)>1$, the contact time during rolling exceeds the initial contact age and interfacial aging begins to dominate over bulk viscoelastic dissipation. \textbf{b}. An array of three blisters are placed following the sequence shown on the image, giving rise to the sequence of peel phase contact ages $t_w^1>t_w^2>t_w^3$. The peel front propagates from left to right. \textbf{c}. Measured force–displacement curve for peeling across the blister array in panel (b). Each blister produces a distinct peel-to-roll transition, resulting in a multistep, heterogeneous adhesion landscape. The magnitude of each force drop is set by the corresponding dwell-time ratio $t_r/t_w$, enabling tunable and sequential adhesion responses in an otherwise homogeneous interface.  }
\label{fig5}
\vspace{-3mm}
\end{figure*}

Observations from Fig.~\ref{fig3} show that rolling initiates when the deformation fields originating at the peel and blister edges are no longer able to remain distinct. This suggests that $\ell_c^r$ is governed by the characteristic decay length over which bending-induced stresses relax within the contact. We therefore estimate this decay length by balancing the deformation of the thin film with the elastic response of the substrate. For a thin film, bending relates local curvature and pressure through the differential relation $p\sim B\kappa''$. If $u$ is the characteristic amplitude of substrate deformation and $\ell$ the horizontal length over which this deformation decays, dimensional arguments yield the scaling $p\sim B u/\ell^4$. In contrast, the elastic substrate responds non-locally to pressure. For an incompressible elastic substrate of thickness $h_a$, the deformation field satisfies, $u(x)=-\int_{-\infty}^{\infty}p(\bar x)\mathcal{K}(x-\bar x)\mathrm{d}\bar x/E_a$, where $\mathcal{K}(x)$ s the Green's function for a localized traction~\cite{johnson1987contact}. The convolution relation between pressure and deformation simplifies in Fourier space, $\tilde{u}(q)=\tilde{p}(q)\tilde{\mathcal{K}}(q)/E_a$, where $q$ is the wave number and $\mathcal{\tilde K}(q)$ is given by ~\cite{sneddon1995fourier}
\begin{equation}
    \mathcal{\tilde K}(q)=\left[\frac{\sinh{(2qh_a)}-2qh_a}{\cosh{(2qh_a)}+2(qh_a)^2+1}\right]{1\over 2q}.
    \label{eq1}
\end{equation} 
In the long wavelength limit $(qh_a)\ll 1$, the above kernel simplifies to, $ \mathcal{\tilde K}(q)\simeq h_a^3q^2/3$, giving $\tilde u(q)=-(h_a^3q^2/(3E_a))\tilde p(q)$, which transforms back to a local differential form in real space,
\begin{equation}
u(x)={h_a^3\over 3E_a}\frac{\partial^2 p(x)}{\partial x^2}.
\label{eq2}
\end{equation} This local form yields a scaling relation for substrate deformation, $u\sim h_a^3p/(3E_a\ell^2)$. Comparing this with $p\sim B u/\ell^4$, we find an expression for the horizontal decay length 
\begin{equation}
\ell\sim\left({B\over E_a}\right)^{1/6} h_a^{1/2}. 
\label{eq3}
\end{equation} Rolling initiates when the contact length becomes comparable to this length scale, $\ell_c\sim\ell$. $\ell$ resembles length scales that have appeared in studies of thin films on elastic foundations~\cite{ghatak2005measuring,dillard2018review}, and is conceptually akin to the decay length in Kaelble’s classic analysis of the debonding zone during peeling~\cite{kaelble1960theory}. It is instructive to note that the onset of rolling does not require overcoming an additional energy barrier and is therefore not controlled by the work of adhesion. Peeling already proceeds at the peel edge, and rolling simply reorganizes the interfacial kinematics through simultaneous reattachment. This reorganization is set by $\ell$, the geometric length scale over which deformation fields interact within the contact. 

To check the validity of the derived scaling relationship (\eqref{eq3}), we compile experimental measurements across all films (marker shapes), bending moduli (marker size), and adhesive thicknesses (color gradient), together with numerical results, as shown in Fig.~\ref{fig4}a. For a given film, the contact length at the onset, $\ell_c^r$ increases with $h_a$ and subsequently saturates; this saturation is more pronounced in the numerical data. Data points for which $h_a>\ell_c^r$ correspond to an effectively unconfined adhesive regime, in which the contact geometry becomes independent of adhesive thickness. Since \eqref{eq3} is derived under the assumption that $h_a\ll\ell$, we focus on the thin-adhesive regime. Accordingly, we rescale the horizontal axis for data satisfying $h_a<\ell_c^r$ in Fig.~\ref{fig4}b. This rescaling reveals a collapse of the measured $\ell_c^r$ onto the predicted length scale $\ell$, spanning more than an order of magnitude variation. The agreement confirms the proposed mechanical interpretation of the peel-to-roll transition and underscores the universality of the identified scaling law. Interestingly, for commercial pressure-sensitive adhesive the scaling law can be further simplified. In these systems the adhesive and film are of equivalent thicknesses ($h_a\simeq h_f$) but differ in modulus by a factor of $10^4 - 10^6$~\cite{villey2015rate}. Substituting these in \eqref{eq3} yields, $\ell\simeq (5-10)h_f$, a decay length of few hundred microns. This estimate quantitatively captures recent experimental observations of the onset of loop shrinking in self-adherent tapes~\cite{wilting2021unloop}.

\noindent\textbf{Emergent dwell time -} Once rolling begins,new adhesive contacts are continuously formed at the blister edge and convected through the contact region before detaching at the peel front. Their lifetime is therefore set kinematically by the rolling contact length and peel speed, defining an intrinsic timescale $t_r = \ell_c^r/V$, the emergent dwell time. This timescale governs how long interfacial bonds can age before failure and thus directly controls the adhesion force during rolling. This behavior contrasts sharply with the peel phase, where peel speed and contact age influence adhesion independently. As shown in the SI, at fixed contact age the peel force scales as $F_p\sim V^{0.55}$ (Fig. S7), where $0.55$ is the rheological exponent indicating that bulk viscoelasticity dissipation dominates the energy release rate~\cite{robbe1998adhesive, barquins1988adherence}. At fixed speed, by contrast, the force increases only weakly with contact age $F_p \sim t_w^{0.14}$ (Fig. S8), reflecting slow interfacial aging~\cite{gent2003adhesion,ghatak2000interfacial,shanahan1991physical}. Rolling couples these two effects: peel speed simultaneously sets both viscoelastic dissipation and the available time for interfacial aging. The discontinuity in force at the peel-to-roll transition thus reflects the abrupt reduction in dwell time, as the interface switches from long-aged during peeling to newly formed and short-lived during rolling.

To test the central role of this emergent dwell time, we replot all our measurements of the force drop $\Delta F$ at the peel-to-roll transition against the dimensionless ratio $\ell_c^r/(V t_w)$ in fig.~\ref{fig5}a, which compares the rolling dwell time to the initial contact age. The data reveal two distinct regimes. For short rolling dwell times $\ell_c^r/V \ll t_w$, the newly formed contacts detach before significant interfacial aging, and $\Delta F$ exhibits a power-law decay with exponent $n \approx 0.55$, matching the viscoelastic exponent of the elastomer (SI Fig.~S1). This correspondence indicates a regime dominated by bulk viscoelastic dissipation. As the rolling dwell time approaches the initial contact age $\ell_c^r/V \sim t_w$, the decay saturates into a plateau, signaling a transition to an interfacial-aging–dominated regime. For even longer dwell times $\ell_c^r/(V t_w) \gg 1$, we anticipate that $\Delta F$ will increase as chain interpenetration and entanglement strengthen the interface, analogous to classic contact-aging behavior in soft adhesives~\cite{shanahan1991physical}.

\section*{Discussion}
Our results show that adhesion in blistered films can be precisely tuned through a single dimensionless parameter: the ratio of contact ages (or dwell times) in the roll and peel phases. This insight enables a direct strategy for programming adhesion responses. A blister array in which the adhered regions ahead of successive blisters have different contact ages (Fig.\ref{fig5}b–c) converts temporal aging into a spatial design principle: each blister undergoes a peel-to-roll transition whose force drop is set by its local dwell-time ratio $t_r/t_w$. Remarkably, even when the film and adhesive layer are identical, simply imposing a spatial sequence of blisters fixes a corresponding hierarchy of contact ages ($(t_{w}^1> t_{w}^2>t_{w}^3)$. As a result, the oldest blister exhibits the largest force drop upon rolling, followed sequentially by newer ones, producing a multistep and heterogeneous adhesion landscape without any material patterning (Fig.\ref{fig5}c). These results demonstrate that the interplay between geometry and kinematics can enable reconfigurable and tunable adhesion profiles.

Finally, this study identifies a new class of interfacial dynamics in thin film–substrate systems governed by curvature-induced traction asymmetry across adhesive contacts. When the zones of influence of a peel edge and a blister edge overlap, the sharply localized curvature at the peel front overwhelms the weaker curvature associated with the blister. The film therefore spontaneously reattaches at the blister edge and transitions into a rolling state. This peel-to-roll transition arises from a coupled balance between film bending, substrate elasticity, and the kinematics of evolving adhesive contacts. Importantly, the underlying mechanism does not rely on peeling as the driving process. Any situation in which two curvature-dominated features\, such as delamination fronts, blisters, kinks, or localized bends, approach within an interaction length is expected to trigger rolling-like migration. For instance, a growing blister encountering a stationary one would generate analogous dynamics, potentially leading to directed motion or blister ripening. More broadly, our results provide a general framework for understanding how curvature-dominated defects interact and reorganize in thin film assemblies, and demonstrate how dynamic adhesion can be engineered through the interplay of geometry, substrate mechanics, and interfacial kinematics.

\section{Materials \& Methods}

\noindent Soft adhesive are prepared using Dow Corning CY52-276, which consists of two components: part A (base polymer) and part B (cross-linker). To prepare the gel, the two components are mixed in a 1:1 ratio and subsequently degassed using a vacuum pump. The degassed mixture is then poured into molds of the desired thickness and cured at room temperature for 48 hours to allow cross-linking. After curing, the samples are left for an additional 72 hours to stabilize before being used in the experiments. The length and width of the adhesive layer are kept constant for all experiments at $240$ mm and 
$29.5$ mm, respectively, while the thickness is varied depending on the experimental requirements. The rheological properties of the elastomer are characterized using a rotational rheometer (ARG2, TA Instruments), and the corresponding data are shown in SI Fig.~S1.

Narrow strips of thin films (width 25.4 mm) are cut using a precision cutter and cleaned with alcohol wipes. Subsequently, the films are buckled and placed on top of the adhesive layer to form the blisters. The height and width of the blister is kept constant at $30$ mm and $50$ mm respectively. We use two different set ups to peel the blistered film off the soft substrate at a given rate and measure the force. First one is  an Instron load frame equipped with a variable-angle peel fixture and load cell, while the second set up is a custom made 90$^{\circ}$ peel device inspired by the recent design ~\cite{gohl2021modular} - the details are shown in Fig. S2. All the experiments were performed under ambient conditions, and images were were recorded using a Nikon D7000 camera.


\section{Acknowledgements}
We acknowledge helpful discussions with Manoj Chaudhury and Jacco Snoeijer. A.K.P. and A.P. acknowledge funding support from the NSF Grant no CMMI-2515674.



\bibliographystyle{ieeetr}
 \bibliography{Blister}

\makeatletter
\let\old@makecaption\@makecaption
\renewcommand{\@makecaption}[2]{\old@makecaption{#1}{#2}}
\makeatother
\captionsetup{format=plain,justification=justified,singlelinecheck=false}

\end{document}